\documentclass[preprint,showpacs,preprintnumbers,amsmath,amssymb] {revtex4}
\usepackage{graphicx}
\usepackage{dcolumn}
\usepackage{bm}
\newcommand{\bee}{\begin{eqnarray}}
\newcommand{\eend}{\end{eqnarray}}
\newcommand{\rmd}{{\rm d}}
\newcommand{\rme}{{\rm e}}
\newcommand{\rmi}{{\rm i}}
\newcommand{\bea}{\begin{eqnarray}}
\newcommand{\eea}{\end{eqnarray}}
\newcommand{\bb}[1]{\mbox{\boldmath$#1$}}

\begin{document}

\title{Real and virtual photons in an external constant
electromagnetic field of most general form}

\author{Anatoly E. Shabad$^1$ and Vladimir V. Usov$^2$}
\affiliation{ $^1$ P.N. Lebedev Physics Institute, Moscow 117924,
Russia\\
$^2$Center for Astrophysics, Weizmann Institute, Rehovot 76100,
Israel}

\begin{abstract}
The  photon behavior in an arbitrary superposition of constant
magnetic and electric fields is considered on most general grounds
basing on the first principles like Lorentz- gauge- charge- and
parity-invariance. We make model- and approximation-independent, but
still rather informative, statements about the behavior that the
requirement of causal propagation prescribes to massive and massless
branches of dispersion curves, and describe the way the
 eigenmodes  are polarized. We find, as a consequence of Hermiticity in the
transparency domain, that adding a smaller electric field to a
strong magnetic field in parallel to the latter causes enhancement
of birefringence. We find the magnetic field produced by a point
electric charge far from it -- a manifestation of magneto-electric
phenomenon. We establish degeneracies of the polarization tensor
that -- under special kinematical conditions -- occur due to
space-time symmetries of the vacuum left after the external field is
imposed.

\end{abstract}

\pacs{14.70.Bh, 11.55.Fv, 41.20.Jb, 12.20.Ds}

\maketitle

\section{Introduction}
Consideration  of a field theory in an external field background
 is a longstanding challenging
physical problem, that attracts attention owing  to many other
reasons apart from the fact that the astrophysical reality or laser
and accelerator technology do propose electromagnetic fields, so
strong as to make the nonlinearity of the theory of electromagnetic
interaction actual. The cases of external fields of comparatively
simple configurations that admit analytical solution to the Dirac
equation allow one to go beyond the perturbation theory by including
fields of arbitrary strength and thereby consider extreme
conditions, under which it is uncertain whether well-probed theories
continue to be correct and where even most firmly established
fundamental laws may be questioned. For this reason we believe that
when approaching such problems it is important to be aware of the
status of various encountered facts by clearly distinguishing, which
of them reflect basic principles and which depend on an
approximation or a model exploited in establishing them.

In this paper we concentrate on an important problem belonging to
this class, namely the behavior of real and virtual electromagnetic
excitations of the vacuum filled with  constant and homogeneous
electric ($\bf E$) and magnetic ($\bf B$)  fields that are
superposed in such a way that the both field invariants
$\mathfrak{F}=({\bf B^2-E^2})/2$ and $\mathfrak{G}=({\bf B\cdot E})$
are different from zero. This makes the most general case of a
constant in space and in time electromagnetic field, such that
neither electric nor magnetic component can be eliminated from it by
a Lorentz transformation, but a special reference frame always
exists, where they are parallel or antiparallel. We shall avoid
calling all these excitations \textit{photons}, even when they are
real and not virtual, because photons only make a massless subclass
of all possible real excitations of the vacuum possessing the same
quantum numbers as photons and supplying poles to the same photon
Green function.

In principle, the excitations  are completely described using the
effective Lagrangian, which, when differentiated upon the fields,
generates all two- and many-excitation vertices. However, as far as
calculations are concerned, the use of the effective Lagrangian is
efficient only if its dependence on the space- and time-derivatives
of the field strengthes is not taken into account. The results
obtained in this way relate to the vanishing excitation momentum
components $k_\mu\rightarrow0,$ $\mu=1,2,3,0,$ and correspond, if
extended beyond this infrared limit, to dispersion curves that are
straight lines passing through the origin in the plane of the
natural kinematical variables (we refer to the special frame)
$k_0^2-k_3^2$, $\bf k_\perp^2,$ where $k_0$ is the energy of the
excitation, and $k_3$ and $\bf k_\perp$ are its momentum projections
onto the common direction (chosen as axis \textbf{3}) of $\bf E$ and
$\bf B$ and onto the transverse plane, respectively. All massive
excitations, i.e. the ones  with nonzero rest energy
$\left.k_0\right|_{{\bf k}=0}\neq0,$ are lost in this limit. Within
this framework the problem of photon propagation in a constant
background, where $\mathfrak{F}$ and $\mathfrak{G}$ are both
nonzero, was first studied by Pleba\'{n}ski \cite{plebanski}, who
appealed to a rather general nonlinear Lagrangian
 independent of the
field-derivatives. Among other results, he wrote certain conditions,
which the Lagrangian should satisfy lest it might contradict the
causal propagation of small-amplitude electromagnetic waves. More
recently, Novello \emph{et al.} \cite{novello} observed an
interesting possibility that in the infrared limit the constant
background field may be geometrized to be represented by an
equivalent metric. The present authors, too, payed attention to the
infrared limit by proving recently \cite{arXiv} the convexity
property of the Lagrangian as a function of the both variables
$\mathfrak{F}$ and $\mathfrak{G}$ in the point $\mathfrak{G}=0,$
basing on causality and unitarity requirements.

To  adequately consider  the linearized problem of small vacuum
excitations the knowledge of the infrared limit is not sufficient,
and one is obliged to appeal to the second-rank polarization tensor
as a function of arbitrary 4-momentum $k_\mu$ not restricted to any
mass shell, i.e. with arbitrary virtuality $k^2$. (To consider the
problem beyond the linear approximation higher-rank polarization
tensors should be taken also at arbitrary values of all the four
momentum components). The needed polarization operator was first
studied by Batalin and Shabad \cite{batalin} (see also the book
\cite{shabtrudy}), who found the general covariant structure and
eigenvector expansion (the diagonal form) of the polarization
operator and photon Green function in the constant field with the
both invariants $\mathfrak{F}$ and $\mathfrak{G}$, taken nonzero
simultaneously, that follows exclusively from the Lorentz- gauge-
and charge-invariance  and parity conservation of Quantum
Electrodynamics (QED). They also calculated the polarization
operator as an electron-positron loop in the external field of
arbitrary strength. Such one-loop calculations were repeated by
Bayer \emph{et al.} and Urrutia \cite{bayer}. The latter author also
studied in more detail the useful further approximation of small
external field and zero virtuality, and observed some features,
which, as a matter of fact, are independent of this approximation,
as well as of the one-loop approximation itself. One-loop
polarization operator  was revisited in \cite{gies} and, under
simplifying kinematical conditions, was separately calculated in
\cite{daugherty}.

The ardor of investigators towards the study of light propagation in
the field that contains electric component in any Lorentz frame was
damped by the fact that within the perturbation theory such field
is, according to Schwinger \cite{schwinger}, unstable with regard to
spontaneous electron-positron pair production. (It is   seen from
\cite{ritus} that the imaginary part of the effective Lagrangian on
constant fields calculated within one- and two-loop accuracy is
nonzero if either $\mathfrak{F}<0, \mathfrak{G}=0$ or
$\mathfrak{F}\lessgtr0, \mathfrak{G}\neq0$). A special theory for
handling such fields was developed in \cite{gitman}. After exploited
in one-loop calculations of the polarization operator in
\cite{shwartsman} it indicated, with the help of the general
analysis in \cite{perez}, that C-invariance (the Furry theorem) is
violated, while PT-invariance (the Onsager theorem) is preserved,
hence there is no CPT. Actually, a matrix (namely,
$\Psi^{(5)}_{\mu\nu}$ in \cite{shwartsman}, \cite{perez}) appears in
the decomposition of the one-loop polarization operator, which  is
odd in the external field and hence incompatible with the
C-invariance, but compatible with the Onsager theorem (see Refs.
\cite{perez},\cite{shabtrudy} for its formulation fit for the
present use), because it is antisymmetric with respect to the
indices. On the other hand, the matrix $\Psi^{(7)}_{\mu\nu}$, also
odd in the external field, but symmetric under the permutation of
its indices, that might contradict also the Onsager theorem, is not
involved. Once the external field instability that contradicts
CPT-invariance and is itself a violation of general principles
depends on the perturbation theory and is unknown beyond it, it is
ignored in the analysis of the present paper based exclusively on
these principles.

During the years that followed most efforts were devoted to
important special case of one-invariant field with $\mathfrak{G}=0$
and $\mathfrak{F}\geq0$ that corresponds to a purely magnetic field
in a special Lorentz frame. We shall refer to it as "magnetic-like".
The general analysis of Ref. \cite{batalin} is also valid in this
case, whereas the corresponding one-loop polarization operator is
contained in the formulae of that work as a simple limit
$\mathfrak{G}=0,$ analyzed specifically a bit later in
\cite{annphys}. This limiting polarization operator was recalculated
separately by Tsai \cite{tsai}. It must be pointed that the one-loop
polarization operator in a magnetic-like field for  vanishing
virtuality, $k^2=0$, had been known earlier, after the important
papers by Adler \emph{et al.} \cite{adler}. The simplification
$k^2=0$ is sufficient for considering  small dispersion  and has
been permanently playing a significant role in astrophysical
applications. It does not serve, however, the case when large
deviations from the vacuum dispersion law take place, as is the case
when cyclotron resonances of the vacuum polarization at the
thresholds of creations of free \cite{annphys} or mutually bound
\cite{ass} - \cite{wunner} electron-positron pairs are exploited to
produce the photon capture \cite{nature} by a strong magnetic field
of pulsars. Besides, the assumption that  $k^2=0$ 
completely excludes massive states and, moreover, the whole of the
one of the three polarization modes that cannot carry massless
excitations.

The reason why the magnetic field attracted so much attention was,
of course, the discovery of extremely strong magnetic fields (up to
$\sim 10^{14}- 10^{15}$ G) in the vicinity of many compact
astronomical objects (soft gamma-ray repeaters, anomalous X-ray
pulsars, and some radio pulsars) identified with rotating neutron
stars \cite{TD95}. Still stronger magnetic fields ($B\sim
10^{16}-10^{17}$~G) were predicted to exist at the surface of
cosmological gamma-ray bursters if they are rotation-powered neutron
stars similar to radio pulsars \cite{U92}. The remarkable feature of
these magnetic field strengths is that they are higher than the
characteristic field value $B_0=m_\rme^2c^3/e\hbar\simeq 4.4\times
10^{13}$ G above which the nonlinearity of QED becomes actual, where
$m_\rme$ and $e$ are the electron mass and its charge, respectively.
[Henceforth, we set $\hbar=c=1$ and refer to the Heaviside-Lorentz
system of units.] The compact astronomical objects identified with
strongly magnetized neutron stars are powerful sources of
electromagnetic radiation, and therefore, propagation of photons in
a strong magnetic field is one of the central problems in the theory
of these objects. Correspondingly, these problems have been
extensively studied aiming at applications to the pulsar physics
(for a review, see \cite{HL06}). In parallel, some more academic
features of nonlinear electrodynamics in a magnetic field were
clarified, such as the linear growth of dielectric constant with the
magnetic field \cite{skobelev}, dimensional reduction of the Coulomb
field of a point source \cite{prl2007}, and the upper bound to the
magnetic field due to positronium collapse \cite{prl}. Also the
notion of anomalous magnetic moment of the photon was introduced
\cite{peres}, especially interesting in the large-field limit owing
to the above linear growth \cite{selym}.

In the meanwhile little attention has been paid to the admixture of
electric field. One of the reasons was that although the electric
field is generated along the magnetic field lines in the
magnetospheres of rotating, strongly magnetized neutron stars
\cite{RS75}, the component $E_\parallel$ in the vicinity of  all
compact astronomical objects mentioned above is sufficiently small
($E_\parallel/B\ll 1$) \cite{RS75} as compared to the magnetic
field. Hence, at the first sight, $E_\parallel$ could result in only
minor corrections \cite{zheng}. However, it may be not the case at
least for some processes that are forbidden without the  electric
field $(E_\parallel =0)$ and might be allowed at $E_\parallel\neq
0$. Splitting of photons in a strong magnetic field $(\gamma
+B\rightarrow \gamma'+\gamma"+B)$ is one of the candidates to be
such a process. The point is that in a  magnetic-like field
splitting of one photon mode is allowed, while splitting of the
other is strictly forbidden \cite{adler,Usov2002}. As far as we are
aware, the polarization selection rules for photon splitting have
never been satisfactorily considered in the case of $E_\parallel\neq
0$. Also it is not  clear beforehand how will the processes that
depend on the resonance behave under inclusion of even small
electric field.

On the other hand, extremely strong electric fields with the
strength as high as $\sim (10\div10^2) E_0$ are predicted to exist
at the surface of bare strange stars that are entirely made of
deconfined quarks, where $E_0=m_e^2/e\simeq 1.3\times 10^{16}$ V/cm
is the characteristic electric field value \cite{AFO86}. These
electric fields are directed perpendicular to the stellar surface
and prevent ultra-relativistic electrons of quark matter from their
escape to infinity. The surface magnetic fields are expected to be
more or less the same for neutron and strange stars (from $\lesssim
10^9$ G to $\sim 10^{15}$ G or even higher), and therefore, at the
surface of strange stars the ratio $E_\parallel/B$ may be both $\ll
1$ and $\gg 1$, i.e., it may be practically arbitrary.

We, therefore, feel that the time has come to renew the study of the
general combination of constant electric and magnetic field as a
strong external background  to smaller electromagnetic fields  on
the base of the recent-years experience acquired  when considering
the magnetic-like field.

In this paper we are elaborating consequences of the general
structure of the polarization operator, established in
\cite{batalin}, for the  propagation and polarization of photons and
massive vector excitations of the vacuum in the presence of a
constant background field with the both field invariants different
from zero. Also some observations are made depending on the one-loop
approximation. These are exiled to Appendix 2.

In Section II we  review  properties of the polarization operator
and of its three excitation eigenmodes that follow from its diagonal
representation, derived in \cite{batalin} from the fundamental
principles in an approximation- and model-independent way. We
present a kinematical orthogonal basis and the decomposition of the
eigenvectors of the polarization operator over it. The  coefficients
in this decomposition are expressed in terms of four linear
invariant combinations, called $\Lambda_{1,2,4,5}$, of the
polarization operator components. We also present the simpler form
of this decomposition valid in the limit, where the mixing between
the basic vectors is small, like for small admixture of an electric
field to a large magnetic one, and under special kinematical
conditions. We establish that under this admixture the distance
between dispersion curves of two modes that can carry massless
excitations (photons) increases in the transparency domain as a
consequence of Hermiticity, thus leading to increasing the
birefringence and strengthening Adler's kinematical bans
\cite{adler} for photon splitting in a magnetic field. We discuss
how Adler's CP-selection rules are modified in the case of the
general field under consideration. We find in the infrared limit the
functions $\Lambda_{1,2,3,4}$, on which the three polarization
operator eigenvalues and eigenvectors depend in an irrational way,
in terms of the field derivatives of the effective action functional
defined on constant background field to see explicitly that the
eigenvalues disappear in the zero point of the excitation 4-momentum
as a consequence of gauge invariance.

In Section III this property is used to ground the statement that
there always exist massless excitations to be identified with
photons present in two polarization modes, whereas any number of
massive branches may be present in all the three modes. The
corresponding excitations have the same quantum numbers as photons
and supply poles to the same propagator. To avoid a possible
misunderstanding it is worth stressing that we do not keep to the
definition of the photon mass as its 4-momentum squared $k^2\neq0$,
used by some authors. In our opinion, such word usage is
counterproductive, because, if followed, it would make us recognize
a photon in a standard isotropic medium with a constant refraction
index as massive. This would make the notion of the photon mass
indiscriminate. On the contrary, for us, the mass of an excitation
is its rest energy, i.e. the value of its frequency (energy) when
its spatial momentum is zero. Then, the excitations, whose
dispersion curves include the origin in the momentum space, are
referred to as photons.

By restricting the group velocity of an excitation below the speed
of light we establish that in the special frame each dispersion
curve is limited from above in the plane ($\sqrt{k_0^2-k_3^2}, |{\bf
k_\perp}|$) by a straight line that crosses the dispersion curve at
$|{\bf k_\perp}|=0$ and is inclined to the coordinate axes at the
angle of $45^{\rm o}$  (see Fig.1). Massless branches are restricted
to the exterior of
 the light cone $\sqrt{k_0^2-k_3^2}\leq |{\bf k_\perp}|$, whereas
massive ones may cross it. This situation is similar to the magnetic
field alone \cite{arXiv}, because the general field also specializes
only one direction in the space, namely the common direction of the
magnetic and electric fields in the special reference frame.

In Section IV we describe, in the special frame, polarizations of
electric and magnetic fields of the eigenmodes and find, specially,
that the electric fields of modes 2 and 3, which are responsible for
massless excitations, lie in a common plane spanned by  two
3-vectors, one of which is orthogonal to the plane, where the
external fields and the propagation momentum $\bf k$ lie, while the
other belongs to that plane and makes a universal angle that depends
only on momentum components with the external fields. Electric field
of mode 1 is polarized along $\bf k_\perp$. In the same section we
find the large-distance behavior of the magnetic field produced by a
point electric charge placed in  the background electromagnetic
field.

In Section V we establish relations to be obeyed by the four
invariant combinations of the polarization operator components
$\Lambda_{1,2,4,5}$ under special kinematical conditions that let
the symmetry of the external field under rotations around its
direction in the special frame and under Lorentz boost along this
direction manifest itself as degeneracies of the polarization
tensor, i.e. coincidences between pairs of its eigenvalues.

Appendix 1 is technical. In Appendix 2 we present the one-loop
approximation for the function $\Lambda_3$ in the limit of small
electric admixture to the external magnetic field,
$\mathfrak{G}\rightarrow0$, responsible for mixing eigenmodes in
this limit. It has cyclotron resonances starting with the second
threshold of electron-positron pair creation by a photon in a
magnetic field. Therefore, the mixing does not affect the photon
capture effect at the first threshold important for radiation
formation in the pulsar magnetosphere.

\section{Polarization operator, its eigenvectors and eigenvalues}
Before starting,  technical conventions are in order. There are two
field invariants  $\mathfrak{F}=\frac
1{4}F_{\rho\sigma}F_{\rho\sigma}$ and $\mathfrak{G}=\frac
1{4}F_{\rho\sigma}\tilde{F}_{\rho\sigma}$ of the background fields
 and  two Lorentz-scalar combinations  $kF^2k$ and
$k\tilde{F}^2k$ of the background field strength and momentum
$k_\mu$ of the elementary excitation, subject to the relation
\bee\frac{k\tilde{F}^2k}{2\mathfrak{F}}-k^2=
\frac{k{F}^2k}{2\mathfrak{F}}\,.\eend The dual field tensor is
defined as $\tilde{F}_{\rho\sigma}=\frac
\rmi{2}{\epsilon}_{\rho\sigma\lambda\varkappa}F_{\lambda\varkappa},$
where the completely antisymmetric unit tensor is fixed in such a
way that ${\epsilon}_{1234}=1$. We use the notations
 $(\tilde{F}k)_\mu\equiv \tilde{F}_{\mu\tau}k_\tau$, $(Fk)_\mu\equiv
F_{\mu\tau}k_\tau$, $F^2_{\mu\nu}\equiv F_{\mu\tau}F_{\tau\nu}$,
$(F^2k)_\mu\equiv F^2_{\mu\tau}k_\tau$, $kF^2k\equiv k_\mu
F^2_{\mu\tau}k_\tau$, $k^2\equiv {\bf k}^2+k_4^2={\bf k}^2-k_0^2$
and are working in Euclidian metrics with the results analytically
continued to Minkowsky space, hence we do not distinguish  co- and
contravariant indices. The scalar variables
 \bee\label{variables}\mathcal{B}=
\sqrt{\mathfrak{F}+\sqrt{\mathfrak{F}^2+\mathfrak{G}^2}},\qquad
\mathcal{E}=\sqrt{-\mathfrak{F}+\sqrt{\mathfrak{F}^2+\mathfrak{G}^2}}\eend
make the meaning, respectively, of the  magnetic,
$\mathcal{B}=B=|\bf B|,$ and electric fields, $\mathcal{E}=E=|\bf
E|$,  in the (special) Lorentz frame, where $\bf B$ and $\bf E$  are
directed along the same axis  chosen as axis \textbf{3} in what
follows. The designation $"\Leftrightarrow"$ will
 establish correspondence between quantities relating to  the general
  Lorentz frame and the values these take in the special frame.
  Referring to the fact that in the special frame
 the momentum-containing invariants become
\bee\label{special}k\tilde
{F}^2k={B}^2(k_3^2-k_0^2)-E^2k_\perp^2,\qquad
{k{F}^2k}=-{B}^2k_\perp^2+E^2(k_3^2-k_0^2),\eend with the
two-dimensional vector ${\bf k}_\perp$ being the  momentum
projection onto the plane orthogonal to the common direction
\textbf{3} of the electric and magnetic field we shall use the
equivalence relations \bee\label{general}
\frac{k^2\mathcal{B}^2+kF^2k}{\mathcal{B}^2+\mathcal{E}^2}\Leftrightarrow
k_3^2-k_0^2, \quad
\frac{k^2\mathcal{E}^2-kF^2k}{\mathcal{B}^2+\mathcal{E}^2}\Leftrightarrow
k_\perp^2 \eend  throughout the paper.

 Polarization operator $\Pi_{\mu\nu}(x,y)$ is responsible for  small perturbations
 above the constant-field background.
It follows from the translation- Lorentz-, gauge-, PT- and
charge-invariance \cite{batalin,shabtrudy} that its Fourier
transform can be presented in a diagonal form \bee\label{pimunu}
\Pi_{\mu\tau}(k,p)=\delta(k-p)\Pi_{\mu\tau}(k),\qquad
\Pi_{\mu\tau}(k)=\sum_{a=1}^3\varkappa_a~\frac{\flat_\mu^{(a)}~
\flat_\tau^{(a)}}{(\flat^{(a)})^2},\eend  where $\flat_\tau^{(a)}$
are its eigenvectors
\begin{eqnarray}\label{eigen}\Pi_{\mu\tau}~\flat^{(a)}_\tau=\varkappa_a~\flat^{(a)}_\mu,\quad
a=1,2,3,4,
\end{eqnarray} while the eigenvalues $\varkappa_a$ are scalar
functions of $\mathfrak{F},$
 $\mathfrak{G},$  $kF^2k$ and
$k\tilde{F}^2k.$

The fourth eigenvector  is trivial, $\flat^{(4)}_\mu=k_\mu$, so the
fourth eigenvalue vanishes, $\varkappa_4=0,$ as a consequence of the
4-transverseness of the polarization operator,
$\Pi_{\mu\tau}k_\tau=0$. All eigenvectors are mutually orthogonal,
$\flat^{(a)}_\mu \flat_\mu^{(b)}\sim \delta_{ab}$, this means that
the first three ones are 4-transversal, $\flat^{(a)}_\mu k_\mu=0$.

In the special case, where the second field invariant disappears,
$\mathfrak{G}=0,$ the three meaningful eigenvectors
$\flat^{(1,2,3)}_\mu$ are known \cite{batalin,annphys, shabtrudy} in
the universal final form:
\begin{eqnarray}\label{vectors}
\left.\flat^{(1)}_\mu\right|_{\mathfrak{G}=0}=(F^2k)_\mu
k^2-k_\mu(kF^2k),\quad
\left.\flat_\mu^{(2)}\right|_{\mathfrak{G}=0}=(\tilde{F}k)_\mu,\quad
\left.\flat_\mu^{(3)}\right|_{\mathfrak{G}=0}=(Fk)_\mu.
\end{eqnarray} This case implies that in the special frame only
magnetic, when $\mathfrak{F}>0$, or only electric, when
$\mathfrak{F}<0,$ field exists. In the limit $k^2=0$ modes 2, 3
 correspond to Adler's \cite{adler} $\perp$- and $\parallel$-modes,
 respectively, whereas mode 1 becomes pure gauge. Vectors
(\ref{vectors}) may be used as a convenient orthogonal basis also
when no external field is present. But when $\mathfrak{G}\neq0,$
they no longer diagonalize the polarization operator already because
the vectors $(\tilde{F}k)_\mu$ and $(Fk)_\mu$  stop being mutually
orthogonal, since their scalar product
$-k\tilde{F}Fk=\mathfrak{G}k^2$ is now nonzero.

When $\mathfrak{G}\neq0$, the first eigenvector is expressed in
terms of the fields by the same formula as in (\ref{vectors}):
\bee\label{first}\flat^{(1)}_\mu=(F^2k)_\mu k^2-k_\mu(kF^2k),\qquad
(\flat^{(1)}\tilde{F}k)=(\flat^{(1)}{F}k)=0, \nonumber\\
(\flat^{(1)})^2=k^2(k^2\mathcal{E}^2-kF^2k)(k^2\mathcal{B}^2+kF^2k)\Leftrightarrow
k^2(B^2+E^2)^2k_\perp^2(k_3^2-k_0^2)\eend and the first eigenvalue
is given by the
formula\bee\label{kappafirst}\varkappa_1=\frac{k^2(\mathcal{B}^2+\mathcal{E}^2)}{k^2
\mathcal{B}^2+kF^2k}\Lambda_1\Leftrightarrow\frac{k^2}{k_3^2-k_0^2}\Lambda_1,
 \eend where the scalar function of the fields and momentum $\Lambda_1$ here, as well as  other $\Lambda$'s  below,
 is a
  {\em linear} superposition of the polarization
 tensor components $\Pi_{\mu\nu}$.
 The other two eigenvectors are the linear
combinations\bee\label{b23}
\flat^{(2,3)}_\mu=-2\Lambda_3c_\mu^{-}+\left[\Lambda_2-
\Lambda_4\pm\sqrt{(\Lambda_2-\Lambda_4)^2+4\Lambda^2_3}\right]c_\mu^{+}\eend
(where the square root is understood algebraically: $\sqrt{Z^2}=Z$,
and not $|Z|$) of  two  orthonormalized vectors
:\bee\label{d}c_\mu^{-}=
\frac{\mathcal{B}(Fk)_\mu+\mathcal{E}(\tilde{F}k)_\mu}
{(\mathcal{B}^2+\mathcal{E}^2)^{1/2}(k^2\mathcal{E}^2-kF^2k)^{1/2}}\Leftrightarrow
\frac{{B}(Fk)_\mu+{E}(\tilde{F}k)_\mu} {({B}^2+{E}^2)|\bf k_\perp
|},
\nonumber\\
c_\mu^{+}=
\rmi\frac{\mathcal{E}(Fk)_\mu-\mathcal{B}(\tilde{F}k)_\mu}
{(\mathcal{B}^2+\mathcal{E}^2)^{1/2}(k^2\mathcal{B}^2+kF^2k)^{1/2}}\Leftrightarrow
\frac{E(Fk)_\mu-B(\tilde{F}k)_\mu}
{(B^2+E^2)(k_0^2-k_3^2)^{1/2}},\nonumber\\\nonumber\\
(c^{+}c^{-})=(c^\pm\flat^{(1)})=(c^\pm k)=0,\quad
(c^\pm)^2=1,\hspace{7cm} \eend thereby of  the former basic vectors
$(\tilde{F}k)_\mu$ and $(Fk)_\mu,$ too. The corresponding two
eigenvalues are
\bee\label{kappa23}\varkappa_{2,3}=\frac1{2}\left[-(\Lambda_2+
\Lambda_4)\pm\sqrt{(\Lambda_2-\Lambda_4)^2+4\Lambda^2_3}\right].\eend
 The scalar coefficients in the linear combination (\ref{b23}) cannot be
expressed in a universal way in terms of the field and momentum, but
are irrational functions of the polarization tensor components. The
reason is that the polarization operator is a linear combination of
four independent matrices with four scalar coefficients, whereas
there may be only three eigenvalues in accordance with three
polarization degrees of freedom of a vector field. (When
$\mathfrak{G}=0,$ the number of independent matrices reduces to
three). The orthogonality $(b^{(2)}b^{(3)})=0$ is explicit in
(\ref{b23}). The Lorentz-invariant coefficients $\Lambda_{1,2,3,4}$
are functions of the background fields and momenta. Expressions for
 them as simple linear superpositions of the components
$\Pi_{\mu\nu},$
\bee\label{sandwich}\Lambda_1=\frac{(kF^2)_\mu\Pi_{\mu\nu}(
F^2k)_\nu}{(\mathcal{B}^2+\mathcal{E}^2)(k^2\mathcal{E}^2-kF^2k)},\quad
\Lambda_2=- c^{-}_\mu\Pi_{\mu\nu}c_\nu^{-},\quad \Lambda_3=-
{c^{-}_\mu\Pi_{\mu\nu}c_\nu^{+}},\quad \Lambda_4=-
{c^{+}_\mu\Pi_{\mu\nu}c_\nu^{+}}\quad\eend are obtained in Appendix
1 from a less transparent representation to be found in
\cite{batalin, shabtrudy};
 their calculations in one-loop approximation of QED are given in
\cite{batalin, shabtrudy}.

The transparency domain of momenta is such a region where absorption
is absent. The electron-positron pair production by a photon is an
example of absorption. The region, where it is kinematically
allowed, is not the transparency domain. The absence of absorption
of small perturbation of the background field is reflected in the
property of Hermiticity \cite{perez} of the matrix $\Pi_{\mu\nu}$.
It is symmetric when the charge conjugation invariance holds
\cite{perez, shabtrudy} (no charge-asymmetric plasma background, no
spontaneous pair creation). Hence, in the transparency region all
the components of $\Pi_{\mu\nu}$ are real in the case under
consideration, once the charge conjugation invariance is assumed.
Then, all $\Lambda$'s defined by (\ref{sandwich}) are also real
there, except the region $k_3^2-k_0^2>0$ (or, in invariant terms,
$k^2\mathcal{B}^2+kF^2k>0$) wherein $\Lambda_3$ becomes imaginary
due to (\ref{d}). (We shall see later that dispersion curves cannot
get into this region without violating the stability). In this
exceptional region the quantity under the square root in
(\ref{b23}), (\ref{kappa23}) stops being manifestly positive.
Nevertheless, it should remain nonnegative, since eigenvalues of a
Hermitian matrix should be real.

The dispersion equations that define the mass shells of the three
eigenmodes are
\bee\label{dispersion}\varkappa_a(k\tilde{F}^2k,kF^2k,\mathfrak{F},\mathfrak{G}^2)=k^2,\qquad
a=1,2,3. \eend We have explicitly indicated  here that the
eigenvalues should be even functions of the pseudoscalar
$\mathfrak{G}$.

When, due to a certain reason, $\Lambda_3$ is small as compared to
$|\Lambda_2-\Lambda_4|,$ the small mixing of eigenmodes is obtained
by expanding (\ref{b23}) in powers of
$\Lambda_3/|\Lambda_2-\Lambda_4|$. In this way we get, with the
linear accuracy in $\Lambda_3$, after normalizing out the common
factors 2 and $2\Lambda_3/(\Lambda_2-\Lambda_4)$
\bee\label{be}\flat_\mu^{(2)}=-\Lambda_3c^-_\mu+(\Lambda_2-\Lambda_4)c^+_\mu,\qquad
\flat_\mu^{(3)}=(\Lambda_2-\Lambda_4)c^-_\mu+ \Lambda_3c^+_\mu.
\eend Such situation occurs, first of all, when the electric field
is small as compared to the magnetic one, which we shall discuss
now, and also for two cases of special kinematical conditions
considered in Section V.

In the limiting regime of small $\mathfrak{G}\rightarrow0$, one has
$\mathcal{E}\approx\mathfrak{G}/\mathcal{B}$. So $\mathcal{E}$ is a
pseudoscalar, hence $c^-_\mu$ is a vector, and $c^+_\mu$ a
pseudovector, the same as $({F}k)_\mu$ and $(\tilde{F}k)_\mu$,
respectively, are. Then, from (\ref{b23}) it follows that
$\Lambda_3$ is a pseudoscalar vanishing linearly:
$\Lambda_3\sim\mathfrak{G}\approx(\mathcal{E}\mathcal{B})\rightarrow0.$
(This fact is also in agreement with the infrared limit (\ref{ir5})
below, since $\mathfrak{L}$ depends on $\mathfrak{G}^2$, and with
the one-loop result in \cite{batalin, shabtrudy}). The eigenvector
$\flat^{(2)}_\mu$ given by eq. (\ref{b23}) with the upper sign in
front of the square root becomes in this limit $c^+_\mu\sim
\tilde{F}k_\mu,$ as prescribed by (\ref{vectors}). On the contrary,
the eigenvector $\flat^{(3)}_\mu$ given by eq. (\ref{b23}) with the
lower sign becomes $c^-_\mu\sim {F}k_\mu,$ because the coefficient
in front of $c^+_\mu$ in (\ref{b23}) decreases as $\Lambda_3^2$. The
coefficient $\Lambda_3$ becomes responsible for mixing  eigenmodes,
characteristic of an external magnetic field due to the perturbation
caused by electric field. (See Appendix 2 for the linearly vanishing
$\mathfrak{G}\rightarrow0$ limit of $\Lambda_3$ as calculated within
one-loop approximation of quantum electrodynamics.)

Bearing in mind that $\Lambda_{2,4}$ are scalars and may, thus,
contain the pseudoscalar $\mathfrak{G}$ only in even powers, from
(\ref{kappa23}) it may be concluded, prior to any dynamical
calculations, that - as long as  external electric field is small as
compared to  magnetic one - the {\em birefringence}, inherent to the
problem of the light propagation when, in the special Lorentz frame,
a single - magnetic, $\mathfrak{F}>0,$ or electric,
$\mathfrak{F}<0$, field is present, is enhanced as soon as the other
field is added in parallel to the already existing
one:\bee\label{birefringence}
|\varkappa_2-\varkappa_3|=\left|\sqrt{(\Lambda_2-\Lambda_4)^2+4\Lambda^2_3}\right|\geq
|\Lambda_2-\Lambda_4|,\eend i.e. the dispersion curves of modes 2, 3
tend to repulse from  each other \cite{footnote3}, as far as they
lie (we shall discuss later why they are to) in the domain
$k_0^2>k_3^2$, where $\Lambda_3$ is real. Thereby, Adler's
kinematical selection rules \cite{adler} that ban some processes of
one photon splitting into two in a magnetic field are strengthened
if an electric field is added. As for his CP-selection rules
\cite{reservation}, those now should be applied to the eigenwaves,
given as (\ref{b23}), and read as follows: among the three
$\gamma$-states involved into the reaction
$\gamma\rightarrow\gamma\gamma$ there may be only two or none of
mode-2 states, since $\flat_\mu^{(2)}$ is a pseudovector, while
$\flat_\mu^{(1,3)}$ are vectors. However, any state, prepared as an
eigenstate in the magnetic field alone may decay into two like
states under the perturbation caused by the electric field
disregarding the initial CP-bans, since the electric field
introduces the  pseudoscalar $\mathfrak{G}$.

The infrared limit, $k_\mu\rightarrow0,$ of the polarization
operator is important. To get it, it is sufficient to have at one's
disposal only
 the effective Lagrangian $\mathfrak{L(F,G)},$ from where
 the dependance on the time- and space-derivatives of the field $F_{\mu\nu}$
 is disregarded \cite{iliopoulos}.
In the limit of vanishing momenta the invariant coefficients
$\Lambda_{1,2,3,4}$ are quadratic functions of $k_\mu$ expressed in
terms
 of the (momentum-independent) derivatives $\mathfrak{L_F}=\partial\mathfrak{L/\partial
 F},$ $\mathfrak{L_{FF}}=\partial^2\mathfrak{L/\partial
 F}^2,$ $\mathfrak{L_{GG}}=\partial^2\mathfrak{L/\partial
 G}^2,$ $\mathfrak{L_{FG}}=\partial^2\mathfrak{L/\partial
 F\partial G}$ as
 follows\bee\label{ir1}\left.\Lambda_1\right|_{k_\mu\rightarrow0}=(k_3^2-k_0^2)\mathfrak{L_F},\eend
\bee\label{ir2}
 \left.\Lambda_2\right|_{k_\mu\rightarrow0}=-k^2\mathfrak{L_F}-
 {\bf k}_\perp^2({B}^2\mathfrak{L_{FF}}+{E}^2\mathfrak{L_{GG}}+2\mathfrak{GL_{FG}}),\eend
 \bee\label{ir4}
 \left.\Lambda_4\right|_{k_\mu\rightarrow0}=-k^2\mathfrak{L_F}+(k_3^2-k_0^2)
 ({E}^2\mathfrak{L_{FF}}+{B}^2\mathfrak{L_{GG}}-2\mathfrak{GL_{FG}}),\eend
 \bee\label{ir5}
 \left.\Lambda_3\right|_{k_\mu\rightarrow0}=
 ({\bf k}_\perp^2)^{\frac1{2}}(k_0^2-k_3^2)^{\frac1{2}}\{\mathfrak{L_{FG}}({B}^2+{E}^2)-(\mathfrak{L_{GG}}+\mathfrak{L_{FF}})
 \mathfrak{G}\}. \eend
We have written these formulae referring to the special frame. The
equivalence relations (\ref{general}) allow to immediately restore
their invariant form valid in any frame. Eqs. (\ref{ir1}) --
(\ref{ir5}) are obtained using the definition of the polarization
operator components as the second derivatives with respect to
vector-potentials components (see, {\textit e.g.} \cite {arXiv}).
Insofar as one is interested in the quantities
$\left.\Lambda_i\right|_{k_\mu\rightarrow0}$ up to one-loop accuracy
one should either take the Heisenberg-Euler expression for
$\mathfrak{L}$ here or pass to the infrared limit in the expressions
for $\Lambda_i$ calculated within one-loop approximation in
\cite{batalin}. The two-loop approximation for $\mathfrak{L}$ is
also available \cite{ritus}.

 From eqs. (\ref{ir1}) -- (\ref{ir5}) the vanishing of the
eigenvalues in the zero-momentum point\bea
\left.\varkappa_a\right|_{k=0}=0, \hspace{5mm} a=1,2,3\label{gauge}
\eea follows. This property is, in the end, a consequence of the
gauge invariance that requires that the effective Lagrangian should
depend only on the field strengthes, and not potentials.

\section{Dispersion curves}

In the special frame  dispersion equations (\ref{dispersion}) can be
represented in the form
\bee\label{dispersion2}\varkappa_a(k_0^2-k_3^2,k_\perp^2,B^2,E^2)=k_\perp^2
+k_3^2-k_0^2,\qquad a=1,2,3 \eend and their solutions that express
the energy $k_0$ of the elementary excitation of a given mode $a$ in
terms of its spatial momentum components $k_3,{\bf k}_\perp$ have
the following general structure, provided, in the end, by the
invariance of the external field under rotation around axis
\textbf{3} and the Lorentz boost along this axis,\bea
k_0^2=k_3^2+f_a(k_\perp^2),\hspace{5mm} a=1,2,3\label{law}, \eea
where the dispersion functions $f_a(k_\perp^2)$ certainly depend
also on the external fields.

The causality principle requires that the modulus of the group
velocity, calculated on each mass shell (\ref{law}), be less than or
equal to the speed of light in the free vacuum $c=1$:
\bee\label{group}|\textbf{v}_{\rm gr}|^2=\left(\frac{\partial
k_0}{\partial k_3}\right)^2+\left|\frac{\partial k_0}{\partial
\textbf{k}_\perp}\right|^2=\frac{k_3^2}{k_0^2}+\left|\frac{\textbf{k}_\perp}{k_0}\cdot
f_a^\prime\right|^2
=\frac{k_3^2+k_\perp^2\cdot(f_a^\prime)^2}{k_3^2+f_a(k_\perp^2)}\leq
1,\eend where $f_a^\prime=\rmd f_a(k_\perp^2)/\rmd k_\perp^2$. This
imposes the obligatory condition on the form and location of the
dispersion curves (\ref{law}), i.e. on the function $f_a(k_\perp^2)$
(remind that $k_3^2+f_a(k_\perp^2)\geq 0 $ due to (\ref{law}))
:\bee\label{causality} k_\perp^2\left(\frac{\rmd
f_a(k_\perp^2)}{\rmd k_\perp^2}\right)^2\leq f_a(k_\perp^2)\,.\eend
This inequality requires first of all that $f_a(k_\perp^2)\geq 0$,
hence no branch of any dispersion curve may get into the region
$k_0^2-k_3^2<0$, where $\Lambda_3$, eq. (\ref{sandwich}), becomes
imaginary. If it might, the photon energy $k_0$ would be imaginary
within the momentum interval $0<k_3^2<-f_a(k_\perp^2),$
corresponding to the vacuum excitation exponentially growing in
time.  This sort of ghost would signal the instability of the vacuum
with a background field. Inequality (\ref{causality}) further
requires that \bee\label{rest}\frac{\rmd
f_a^{\frac1{2}}(k_\perp^2)}{\rmd k_\perp}\leq 1,\qquad {\rm or}
\qquad f_a^{\frac1{2}}(k_\perp^2)\leq f_a^{\frac1{2}}(0)+k_\perp,
\eend where $\mathfrak{m}=f^{1/2}(0)$ is the rest energy (mass) of
the elementary excitation:
$\mathfrak{m}^2=\left.(k_0^2-k_3^2)\right|_{k_\perp=0}=\left.k_0^2\right|_{k_3={\bf
k}_\perp=0}.$ The inequality
\bee\label{rest3}(k_0^2-k_3^2)^{1/2}\leq \mathfrak{m}+k_\perp \eend
 that follows from (\ref{rest}) and (\ref{law}) is an obligatory restriction imposed
by causality principle on the dislocation of dispersion curves in
the presence of constant magnetic and electric fields. In the empty
space the restriction that appears in the similar way is $k_0\leq
\left.k_0\right|_{\bf k=0}+|\bf k|.$ It is certainly obeyed by the
free massive particle: $k_0=({\bf k}^2+\mathfrak{m}^2)^{1/2}\leq
\mathfrak{m}+|\bf k|$, where $\mathfrak{m}=\left.(k_0)\right|_{\bf
k=0}.$

 The gauge invariance property (\ref{gauge}) implies via equation
(\ref{dispersion}) that for each mode there always exists a
dispersion curve with $\mathfrak{m}^2=f_a(0)=0$, which passes
through the origin in the $(k_0^2-k_\parallel^2, k_\perp^2)$-plane.
It is such branch that is to be called {\em a photon}, since it is
massless in the sense that the energy $k_0$ turns to zero for the
excitation at rest, $k_3=k_\perp=0$ (although, generally, $k^2\neq
0$ where $\bf k\neq 0$). Other branches for each polarization mode
$a$  may also appear provided that a dynamical model includes  a
massive excitation of the vacuum with quantum numbers of a photon,
for instance the positronium atom \cite{ass,leinson,wunner,
shabtrudy} or a massive (pseudo)scalar particle (axion) in a
gauge-invariant interaction with the electromagnetic field
\cite{gabrielli}. Note that while the number of polarization modes
of a vector particle is three -- in correspondence with the
dimension of the space and, hence, with the number of degrees of
freedom, -- dispersion curve for each of the three modes may have
any number of branches, \textit{e.g.} an infinite number of excited
positronium branches. The energy on a dispersion curve should be
real, since the dispersion equation (\ref{dispersion}) supplies
poles to the photon propagator \cite{batalin, shabtrudy}
\begin{eqnarray}\label{decomposition} D_{\mu\nu}(k)= \sum_{a=1}^3
\frac{\flat_\mu^{(a)}~ \flat_\nu^{(a)}}{(\flat^{(a)})^2}\frac
1{k^2-\varkappa_a(k)},
\end{eqnarray}(defined up to arbitrary longitudinal part $\sim k_\mu
k_\nu$),
and these should not get into complex plane. If the state is
unstable and should therefore decay, its energy must have an
imaginary part, indeed, but in this case the pole is located on an
nonphysical sheet of the complex plane, whose presence must be
provided by branching points in the polarization operator to be
introduced within an approximation where the state is expected to be
unstable. An example of such situation is given by the cyclotron
resonance \cite{annphys} approximation of the polarization operator
in a magnetic field. The corresponding dispersion equations are
cubic with respect to energy squared. Out of its three solutions,
one corresponds to a stable state and, therefore, is real, whereas
the other two mutually complex conjugated branches responsible for
the photon decay/capture to electron-positron pairs belong to
nonphysical sheets of the complex energy plane. Neither of these
solutions can be disregarded. As applied to unstable branches our
appeal to the group velocity may be reasonable only provided that
the absorption is small, i.e. the imaginary part is much less than
the real one.

 On the other hand, the number of
massless modes is, as a matter of fact, not three, but only two, as
it should be for a photon. The point is that the massless branch of
the dispersion curve for mode 1 does not correspond to any real
elementary excitation, except for two special cases. It follows from
(\ref{kappafirst}) and (\ref{ir1}) that
\bee\label{infrared}\left.\varkappa_1\right|_{k_\mu=0}=k^2
\frac{\partial\mathfrak{L}(\mathfrak{F,G})}{\partial\mathfrak{F}}.\eend
Hence the light cone $k^2=0$ is a guaranteed solution to the
dispersion equation (\ref{dispersion}) for mode 1 in the vicinity of
the origin $k_\mu=0$. Can there exist massless excitations in mode 1
other than $k^2=0$? The answer is "no", because from
(\ref{infrared}) it follows that
 $k^2=0$ is the only possibility for a dispersion curve of mode 1,
as it approaches the origin $k_\mu=0$. Now, from eq. (\ref{first})
it is seen,  that both electric and magnetic fields in  mode 1
disappear at $k^2=0$ , since on this mass shell the elementary
excitation is pure gauge, unless either
${k^2\mathcal{E}^2-kF^2k}\sim k_\perp^2 =0$ or
${k^2\mathcal{B}^2+kF^2k}\sim k_3^2-k_0^2=0$, in which cases the
common factor $k^2$ can be normalized out from $\flat_\mu^{(1)}$
(\ref{first}). These exceptional cases  propose kinematical
conditions for degeneracy of the polarization tensor to be discussed
in Section V. For the first of them, the one of parallel
propagation, $k_\perp^2 =0,$ the mode-1 photon is actual, while, on
the contrary, mode-2 photon no longer exists, the mode-2 excitation
becoming massive, as it is argued below in Section V. The overall
number of massless degrees of freedom, therefore, is again two. Note
that although ${\bf k_\perp}=0$ may seem to be an isolated point, as
a matter of fact this is not the case: every nonparallel propagation
${\bf k_\perp}\neq0$ reduces to perpendicular propagation $k_3=0$ by
a Lorentz boost along axis \textbf{3}, which does not lead us out of
the special frame. In the second exceptional case, $k_3^2-k_0^2=0$,
again the mode-1 photon is actual, but the mode-3 photon does not
exist according to Section V. So the number of massless degrees of
freedom is two in this case, too.

We concluded above in this Section that the causality requires that
in the plane $(\sqrt{k_0^2-k_3^2}, k_\perp)$ the photon dispersion
curves ($\mathfrak{m}=0$) are located outside or coincide with the
light cone: $k^2 \geq 0$. (Remind that the light cone $k^2={\bf
k}^2-k_0^2= 0$  is the mass shell of a photon in the vacuum without
an external field.) However, unlike the case, indicated below eq.
(\ref{causality}), a violation of this ban would not lead to a
complex-energy ghost or directly signalize the vacuum instability,
but would mean a presence of superluminal wave, known as tachyon. On
the other hand, massive branches of the dispersion curves as
restricted by the condition (\ref{rest3}) with
$\mathfrak{m}=f^{1/2}(0)\neq0$ may well cross the light cone and
pass to its exterior. They may even quasi-intercept with the
massless (photon) branches or with branches possessing different
$\mathfrak{m}$. The quasi-interceptions, -- i.e. the would-be
interception of dispersion curves of two states taken as independent
within a certain approximation,--  would result in the mutual
repulsion of the dispersion curves leading to formation of mixed
states, polaritons, an example of which is given by a
photo-positronium - a mixed state between a photon and the
electron-positron bound state created by it in a strong magnetic
field \cite{ass,leinson,wunner,shabtrudy}. This situation is
illustrated by Fig.1.
\begin{figure}[htb]
  \begin{center}
   \includegraphics[bb = 0 0 405 210,
    scale=1]{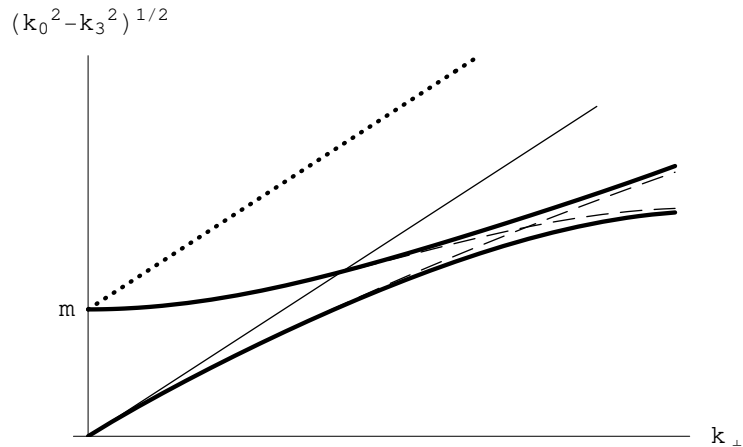} \caption{Disposition of dispersion curves.
    The lower bold solid line is the massless ($photon$) dispersion
    curve restricted from above by the light cone $k^2=0$, presented
    by the thin solid line. The upper solid bold line, representing
    a massive branch, cannot pass higher than the straight dotted line,
    originating from its crossing with the vertical axis according
    eq. (\ref{rest3}). The dashed lines show a quasi-interception.} \label{fig:1}
  \end{center}
  \end{figure}

 The refraction index squared
$n^2_a$ is defined for photons of mode \emph{a} on the mass shell
(\ref{law}) as \bea\label{refrindex} n_a^2\equiv\frac{|{\bf
k}|^2}{k_0^2}
=1+\frac{k_\bot^2-f_a(k_\bot^2)}{k_0^2} .\eend It follows from
 (\ref{rest}) with $\mathfrak{m}=f^{1/2}(0)=0$ that the refraction index is greater
 than unity - the statement common in standard optics of media
 (this is not, certainly, true for (massive) plasmon branches).
Consequently, the modulus of the phase velocity in each mode ${\bf
v}^{\rm ph}_a=(k_0/|\bf k|^2){\bf k}$ equal to $1/n_a$ is, {\em for
the  photon proper}, also smaller than the velocity of light in the
vacuum $c=1$. This is not the case for a massive -- {\em e.g.}
positronium -- branch of the photon dispersion curve, where ${|\bf
v}^{\rm ph}_a|>1$ without any importance for causality.

Now that we established that for photons one has ${\bf k}^2\geq
k_0^2$, or $k^2\geq 0$, we see from the dispersion equation
(\ref{dispersion})  that the eigenvalues $\varkappa_a$ are
nonnegative in the momentum region, where the photon dispersion
curves lie, i.e. the polarization operator is nonnegatively defined
matrix there.
\section{Electromagnetic fields of small perturbations of the
background field}
\subsection{Polarization of eigenmodes}
In the special  frame some peculiarities can be revealed about
orientations of electric and magnetic  fields in  the virtual or
real eigenmodes, formed out of  eigenvectors $\flat^{(a)}_l$ as of
4-potentials according to the standard rules ${ e}^{(a)}_m=\rmi(k_4
\flat^{(a)}_m-k_m\flat^{(a)}_4),$
$h_m^{(a)}=\rmi\varepsilon_{mnl}\flat^{(a)}_nk_l$. To this end let
us write down the eigenvector $\flat_\mu^{(1)}$ (\ref{first}) (after
normalizing it) and the basic vectors $c^{\pm}_\mu$ (\ref{d}), in
the special
frame:\bee\label{bcomp}\frac{\flat_\nu^{(1)}}{\sqrt{|(\flat^{(1)})^2|}}
=\frac{\sqrt
{k_0^2-k_3^2}}{\sqrt{|k^2|}}\left(\begin{tabular}{c}$\frac{\bf
k_\perp}{|\bf k_\perp|}$\\$\frac{k_3|\bf
k_\perp|}{k_0^2-k_3^2}$\\$\frac{-\rmi k_0|\bf
k_\perp|}{k_0^2-k_3^2}$\end{tabular}\right)_\nu,\quad
c_\nu^{-}=\left(\begin{tabular}{c}$\frac{[{\bf
k_\perp}\times{\bb{\epsilon}}]}{|k_\perp|}$\\0\\0\end{tabular}\right)_\nu,\quad
c_\nu^{+}=\left(\begin{tabular}{c}0\\$\frac{-
k_0}{\sqrt{k_0^2-k_3^2}}$
\\$\frac{\rmi k_3}{\sqrt{k_0^2-k_3^2}}$\end{tabular}\right)_\nu.\eend
The upper positions in every column are occupied by two-component
vectors in the perpendicular plane (\textbf{1,2}), next go the third
and fourth components. The normalizing factor is
$|(\flat_\nu^{(1)})^2|= (E^2+B^2)^2(k_0^2-k_3^2)|k^2||{\bf
k_\perp}|^2.$ Here ${\bb{\epsilon}}$ is the unit vector along axis
\textbf{3} and $\textbf{[}{\bf
k_\perp}\times{\bb{\epsilon}}\textbf{]}$ stands for the vector
product: $\textbf{[}~{\bf k_\perp\times{\bb{\epsilon}}
\textbf{]}}_m=\varepsilon_{mns}({\bf k}_\perp)_n\epsilon_s$. The
difference $k_0^2-k_3^2$ is understood to be nonnegative for real
excitations, but may be imaginary for virtual ones.

 From the normalized expression in (\ref{bcomp}) for $\flat^{(1)}_\nu$ we find for the electric and
 magnetic fields in mode 1:  \bee\label{e1}{\bf e}^{(1)}= k_0\frac{{\bf
k_\perp}}{|\bf k_\perp|}\sqrt{\frac{|k^2|}{k_0^2-k_3^2}},\qquad {\bf
h}^{(1)}= \textbf{[}~{{\bf k_\perp}\times{\bb{\epsilon}}
\textbf{]}}\frac{k_3}{|\bf
k_\perp|}\sqrt{\frac{|k^2|}{k_0^2-k_3^2}}.\eend Naturally, the
electric and magnetic fields in mode 1 are oriented in the same way
as  in the single-invariant external field: they both lie in the
plane, orthogonal to the external fields, they are mutually
orthogonal; besides, the magnetic field is transverse, $({\bf
h^{(1)}k})=0$,
 while the electric field, generally, is not: $({\bf e^{(1)}k})\neq0$;
 it is transverse for the special case of  propagation along the external fields $k_\perp=0$.
It is seen again  that a massless excitation is possible in mode 1
as long as it propagates along the external field, otherwise the
fields (\ref{e1}) vanish if $k^2=0$, unless ${\bf k_\perp}=0$, when
the square root turns to unity.

The electric and magnetic fields carried by the vectors
 $(\tilde{F}k)_\mu$ and $({F}k)_\mu$, respectively, are
 \bee\label{eF}{\bf e}^{(\tilde{F})}=-{\bf
k_\perp}{B}k_3+\textbf{[}{\bf k_\perp}\times{\bb
{\epsilon}}\textbf{]}\rmi k_4E-{\bf B}(k_4^2+k_3^2),\quad {\bf
h}^{(\tilde{F})}=-{\bf k}_\perp Ek_3-\textbf{[}{\bf
k_\perp}\times{\bb{\epsilon}}\textbf{]}Hk_0+{\bb
\epsilon}E{k^2_\perp},\qquad\eend \bee\label{eF*}{\bf
e}^{({F})}={\bf k_\perp}{E}k_3+\textbf{[}{\bf
k_\perp}\times{\bb{\epsilon}}\textbf{]}\rmi k_4B+{\bf
E}(k_4^2+k_3^2), \quad{\bf h}^{({F})}=-{\bf k}_\perp
{B}k_3+\textbf{[}{\bf
k_\perp}\times{\bb{\epsilon}}\textbf{]}Ek_0+{\bb{\epsilon}}B{k^2_\perp}.\qquad\eend
The
 electric  fields carried by the vectors
 $c^+_\mu$ and $c^-_\mu$, respectively, are
 \bee\label{e+}{\bf
e}^{+}=\{{\bf k}_\perp k_3+{\bb{\epsilon}}(k_3^2-k_0^2)\}\sqrt{\frac
{(\mathcal{B}^2+\mathcal{E}^2)}{(k^2\mathcal{B}^2+kF^2k)}}={\bf
k_\perp}\frac{k_3}{\sqrt{k_3^2-k_0^2}}+{\bb{\epsilon}}\sqrt{k_3^2-k_0^2},\quad
\quad\nonumber\\({\bf e}^+{\bf e}^-)=0,\quad({\bf
e^{+}k})\neq0,\qquad\eend \bee\label{e-}{\bf e}^{-}
 =\textbf{[}{\bf k_\perp}\times{\bb{\epsilon}}\textbf{]}k_0 \sqrt{\frac
{(\mathcal{B}^2+\mathcal{E}^2)}{(k^2\mathcal{E}^2-kF^2k)}}=\textbf{[}{\bf
k_\perp}\times{\bb{\epsilon}}\textbf{]}\frac{k_0}{|k_\perp|},\quad
({\bf e^-k})=0,\eend
 while their magnetic fields are \bee\label{h-} {\bf
h}^{+}=\textbf{[}{\bf
k}_\perp\times{\bb{\epsilon}}\textbf{]}\frac{k_0}{\sqrt{k_3^2-k_0^2}},\quad{\bf
h}^{-}=-{\bf k_\perp}\frac
{k_3}{k_\perp}+{\bb{\epsilon}}k_\perp,\nonumber\\({\bf h^\pm
k})=({\bf h^+\bf h}^-)=({\bf h^\pm\bf e}^\pm)=0. \eend The
orthogonality of the  electric fields $({\bf e}^+{\bf e}^-)=({\bf
e}^{(1)}{\bf e}^-)=0,$ seen in eqs. (\ref{e1}), (\ref{e+}),
(\ref{e-}), originates from the fact that, in the special frame, the
time-component of one of their mutually orthogonal ($c^+c^-=0$), and
4-transversal ($c^\pm k=0$) vector-potentials $c^\pm_\mu$
disappears: $c^-_0=0$. The 3-vector $\bf e^-$ is directed along the
axis (call it axis \textbf{1}), orthogonal to the plane, where the
external fields and the propagation vector $\bf k$ lie (the plane
\textbf{(3,2)}). The vector $\bf e^+$ lies in that plane. It makes
the universal angle $\alpha=\arctan (k_3k_\perp/(k_0^2-k_3^2))$ with
the direction of the external fields \textbf{3}. Also the magnetic
field ${\bf h}^{-}$ lies in the plane spanned by the external fields
and the propagation direction, while ${\bf h}^{+}$ is orthogonal to
this plane.

The electric fields in the eigenmodes 2 and 3, ${\bf e}^{(2,3)},$
lie both in the common plane spanned by  the vectors ${\bf e}^\pm$:
\bee\label{e23} {\bf e}^{(2,3)}_\mu=-2\Lambda_3{\bf
e}_\mu^{-}+\left[\Lambda_2-
\Lambda_4\pm\sqrt{(\Lambda_2-\Lambda_4)^2+4\Lambda^2_3}\right]{\bf
e}_\mu^{+}.\eend These are not, generally, mutually orthogonal,
since ${\bf e}^\pm$ are not unit-length vectors. Also the magnetic
fields of the modes 2, 3 lie in the common plane spanned by the two
vectors ${\bf h}^{\pm}$ and are linearly combined of them with the
same coefficients as in (\ref{e23}). It can be checked  that the
electric and magnetic fields in each mode are mutually orthogonal,
of course: (${\bf h}^{(2,3)}{\bf e}^{(2,3)})=0$.

In the special case of only one invariant different from zero,
$\Lambda_3\sim\mathfrak{G}\approx(\mathcal{E}/\mathcal{B})\rightarrow0,$
the eigenvectors $\flat^{(2,3)}_\mu$ (\ref{vectors}) are the same as
$c^\pm_\mu$ (\ref{d}) owing to (\ref{b23}), hence ${\bf e}^\pm$,
eqs. (\ref{e+}), (\ref{e-}), and ${\bf h}^\pm$, eq. (\ref{h-}),
become the electric and magnetic fields of the corresponding
eigenmodes, coinciding with their expressions known from Refs.
\cite{batalin, annphys} with the particular property, that the
electric field of mode 2 lies in the plane (\textbf{3,2}), while
that of mode 3 is orthogonal to this plane, known for the special
case of zero virtuality, $k^2=0,$ from Ref. \cite{adler}.
\subsection{Magnetoelectric effect}

In the magnetic-like field it is known that virtual photons of mode
2 are carriers of electrostatic \cite{prl2007, arXiv} forces,
whereas those of modes 1 and 3 are responsible for magnetostatic
interaction \cite{arXiv}. These statements follow from  the
representation (\ref{bcomp}) for the basic vectors, that become
eigenvectors in that special case, and from the diagonal
representation of the Green function (\ref{decomposition}) that
allows to write electric and magnetic fields created by various
(static included) configurations of small -- as compared to the
background field -- charges and currents. The mixing of the basic
vectors (\ref{b23}) in eigenmodes 2 and 3 for the general external
field with $\mathfrak{G}\neq0$ makes these statements no longer true
in what concerns these modes, mode 3 remaining as it was. Moreover,
thanks to the mixing, a  static electric charge, if placed in the
external field with the both invariants different from zero, gives
rise not only to an electric field, as usual, but also to a magnetic
field of its own, like a magnetic charge or moment. Also stationary
currents produce some electric admixture to their customary magnetic
fields.

Here we consider this analogue to the magneto-electric effect known
in crystals \cite{ll} using the field of a point-like static charge
$q$ taken at rest in the special frame as an example. We set the
4-current corresponding to this source in the coordinate space $x$
as $j_\mu(x)=q\delta_{\mu0}\delta^3({\bf x})$, where $\delta_{\mu0}$
is the Kronecker symbol and $\delta^3({\bf x})$ is the Dirac
delta-function. Integrating this current with the Green function
(\ref{decomposition}) we obtain for the vector-potential produced by
the point charge (see \cite{prl2007} for a more detailed explanation
if needed) \begin{eqnarray}\label{potmag}
\hspace{-3cm} A_\mu({\bf x})
=\frac{q}{(2\pi)^3}\int D_{\mu 0}(0,{\bf k})\exp(-\rmi{\bf
kx})\rmd^3k.
\end{eqnarray} Here the argument 0 of the Green function stands for
$k_0$. Among the basic vectors (\ref{bcomp}) there is only one whose
fourth component remains nonzero in the static limit $k_0=0$. It is
$c^+_\mu$. It participates in the eigenvectors $\flat_\mu^{(2,3)}$
in accord with (\ref{b23}). Hence only these two eigenvectors will
remain in the decomposition (\ref{decomposition}) of $D$ after it is
substituted into (\ref{potmag}). On the other hand  the contribution
of the basic vector $c^-_\mu$ may only supply the spacial components
to the vector-potential (\ref{potmag}), whereas $c^+_\mu$ cannot.
Bearing in mind that eqs. (\ref{bcomp}) imply in the static limit,
$k_0=0$, that\bee\label{static}c^+_i=c^-_0=0,\qquad c^+_0=1,\qquad
c^-_i=\frac{\textbf{[}{\bf
k_\perp}\times{\bb{\epsilon}}\textbf{]}_i}{|\textbf{k}_\perp|},\qquad
i=1,2,\qquad c^\pm_3=0\eend the spacial part of the latter is
\bee\label{Aspa} A_i({\bf x})=\frac{q}{(2\pi)^3}\int \sum_{a=2,3}
\frac{\flat_i^{(a)}~ \flat_0^{(a)}}{(\flat^{(a)})^2}\frac
{\exp(-\rmi{\bf kx})\rmd^3k}{{\bf
k}^2-\varkappa_a}=\frac{q}{(2\pi)^3}\int c^-_i\frac
{\Lambda_3\exp(-\rmi{\bf kx})\rmd^3k}{({\bf
k}^2-\varkappa_2)(\textbf{k}^2-\varkappa_3)},\qquad i=1,2\nonumber\\
A_3(\textbf{x})=0. \hspace{14cm}\eend To find the large distance
behavior of this field note that it is determined by the  limit
$\textbf{k}=0$ in the pre-exponential factor in the integrand.
Therefore, we set $\varkappa_{2,3}=0$ and use eq. (\ref{ir5})  for
$\Lambda_3$ \bee\label{ir}
 \left.\Lambda_3\right|_{k_\mu\rightarrow0}=
 \rmi |{\bf k}_\perp|k_3\mathfrak{M},\qquad  \mathfrak{M}= \mathfrak{L_{FG}}({B}^2+{E}^2)-(\mathfrak{L_{GG}}+\mathfrak{L_{FF}})
 \mathfrak{G}. \eend Then
 \bee\label{remote}\left.\textbf{A}(\textbf{x}_\perp,x_3)
 \right|_{|\textbf{x}|\rightarrow\infty}\simeq \frac{q\mathfrak{M}}{(2\pi)^3}\int
\rmi
k_3\textbf{[}\textbf{k}_\perp\times{\bb{\epsilon}}\textbf{]}\frac{\exp(-\rmi{\bf
kx})\rmd^3k}{\textbf{k}^4}. \eend This vector in the two-dimensional
plane orthogonal to the external fields is directed as
$\textbf{[}\textbf{x}_\perp\times{\bb\epsilon}\textbf{]}$, since the
coordinate vector $\textbf{x}_\perp$  in that plane fixes  the only
direction on which the integral may depend. The length of
(\ref{remote})
\bee\label{length}|\textbf{A}|_{|\textbf{x}|\rightarrow\infty}\simeq
\frac{q\mathfrak{M}}{8\pi}\frac1{|\bf x|}. \eend decreases via the
Coulomb law with the radial distance $|\bf x|$ from the charge. The
vector potential and the magnetic field carried by it at large
distances
are\bee\label{Ah}\left.\textbf{A}(\textbf{x}_\perp,x_3)\right|_{|\textbf{x}|\rightarrow\infty}\simeq
\frac{\textbf{[}\textbf{x}_\perp\times{\bb{\epsilon}}\textbf{]}}{|\textbf{x}_\perp|}\frac{q\mathfrak{M}}{8\pi}\frac1{|\bf
x|},\nonumber\\
h_3=\frac{q\mathfrak{M}}{8\pi}\left(\frac1{|\bf
x||\textbf{x}_\perp|}-\frac{|\textbf{x}_\perp|}{|\bf x|^3}\right),
\qquad {\bf h}_\perp=
\frac{q\mathfrak{M}}{8\pi}\frac{\textbf{x}_\perp}{|\textbf{x}_\perp|}\frac
{x_3}{|\bf x|^3}.\eend  The last two equations together make a
magnetic field oriented along the radius-vector (when $q$ is
positive) in the upper half-plane $x_3>0$ and opposite to the
radius-vector in the lower half-plane: \bee\label{magnet}{\bf
h}=\frac{\bf x}{|\bf x|}\frac{q\mathfrak{M}}{8\pi}\frac1{|\bf
x|^2}\frac{x_3}{|\bf x_\perp|}.\eend The magnetic  lines of force
make a pencil of straight lines passing through the origin where the
charge is located and go radially from/to the charge with their
density being the cotangent of the observation angle increasing
towards the axis \textbf{3}, where it is singular. The magneton
$q\mathfrak{M}/(8\pi)$ is proportional to the pseudoscalar
$\mathfrak{G}$; in perturbation theory it has the fine structure
constant $\alpha$ as its overall factor. The result (\ref{magnet})
is approximation-independent, but holds true only as long as the
magnetic field produced by the charge may be considered as a small
perturbation of the background field.

It is worth noting that analogous magneto-electric phenomenon should
be present in a plasma with external magnetic field. The reason is
again in the mixing -- after plasma is added -- of basic vectors
\cite{perez} (see also \cite{shabtrudy}), which are electric and
magnetic carriers in the magnetized vacuum alone. If the plasma is
charge-symmetric, \textit{e.g.} consists of equal numbers of
positively and negatively charged otherwise identical particles, say
electrons and positrons,
 the vector
$\left.\flat^{(1)}_\mu\right|_{\mathfrak{G}=0}$ from (\ref{vectors})
linearly combines with the pseudovector $
\left.\flat_\mu^{(2)}\right|_{\mathfrak{G}=0}=(\tilde{F}k)_\mu$, to
become an eigenvector, unlike the situation considered  above, while
$\quad \left.\flat_\mu^{(3)}\right|_{\mathfrak{G}=0}=(Fk)_\mu$
remains an eigenvector. The pseudoscalar, needed for this
combination, is built of the same pseudovector contracted with the
vector of 4-velocity of the plasma. It plays the role of
$\mathfrak{G}$. For a more general charge-non-symmetric case, say,
electron gas or a gas of ionized atoms, the situation is even more
rich, because all the three basic vectors mix.

\section{Degeneracies}
We have to consider the degeneracies of the polarization matrix
taking place for two special kinematical conditions owing to  the
symmetries of the external field.

For the real or virtual excitations directed parallel to the
external field in the special frame the polarization operator is
symmetric under spacial rotations around the common field direction
\textbf{3}, since the external field is invariant under them, while
the excitation does not introduce an additional anisotropy in the
perpendicular plane owing to the relation $\bf k_\perp=0$. The
symmetry of the polarization operator should manifest itself as a
degeneracy that implies that two out of its three eigenvalues should
coincide, and their corresponding eigenvectors should transform
through one another under the symmetry transformations, while
remaining eigenvectors. By inspecting (\ref{bcomp}) we see that
$\flat_\mu^{(1)}$ and $c^-_\mu$ do possess this mutual property,
when $\bf k_\perp=0$, but $c^+_\mu$ does not (recall that
$|k^2|=k_0^2-k_3^2$, once $\bf k_\perp=0$). Hence the latter cannot
be admixed  to $c^-_\mu$ in (\ref{b23}), and we conclude that, the
same as in the one-invariant external field,
 \bee\label{degen1}\Lambda_3=0,\quad{\rm and}\quad\varkappa_1=\varkappa_3,\quad
{\rm when}\; k_\perp^2=0. \eend (The condition $k_\perp^2=0$ in the
general frame should be replaced by $k^2\mathcal{E}^2-kF^2k=0$).
Then eqs. (\ref{b23}), (\ref{kappa23}) imply
that\bee\label{Lambda12}\Lambda_1=-\Lambda_2,\qquad{\rm when}\quad
k_\perp^2=0.\eend Now from (\ref{e1})--(\ref{h-}) we see that, when
$k_\perp^2=0,$ modes 1 and 3 carry mutually perpendicular and equal
in magnitude transverse electric fields $\bf e^{(1)}$ and  $\bf
e^{(3)}= \bf e^-$ polarized in the plane orthogonal to their
propagation direction and to the external fields. The same is true
for the magnetic fields in these modes. Therefore, in this special
case the mode 1 does correspond to a real photon, as explained in
Section III. Simultaneously, mode-2 becomes a purely longitudinal
wave that may exist only as long as $k_3^2-k_0^2\neq0$, since
(\ref{e+}) otherwise disappears, i.e. it may only be massive. The
conclusion is that the number of massless modes remains
 two, as expected.

 Another degeneracy of polarization operator is provided by the
 kinematical situation $k_0^2-k_3^2=0$. This condition is invariant
 under Lorentz boosts along the common direction of the fields in
 the special frame. (Its invariant equivalent is
 $k^2\mathcal{B}^2+kF^2k=0$). On the other hand, the external field
 in the special frame is also invariant under this transformation:
 it does not lead out of this frame, since the constant electric and magnetic
 fields are not transformed by it. Hence, the polarization operator
 should be also invariant, which implies that some two of its
 eigenvalues must coincide, while the corresponding eigenvectors are
 transformed through one another by the Lorentz rotation in the
 (\textbf{3,0}) hyperplane. This is the case for the vectors
 $\flat_\nu^{(1)}$ and $c^+_\nu$ in (\ref{bcomp}), since in the
 limit under consideration the 2-vector in the upper row of the
 former is negligible as compared to the other two components, and
 $|{\bf k_\perp|}=\sqrt{|k^2|},$ so that $\flat_\nu^{(1)}$ matches
$c^+_\nu$ as a Lorentz boost partner. On the contrary, the vector
$c^-_\nu$ does not transform through any of then, because it is
Lorentz-boost-invariant. We conclude that $c^-_\nu$ and $c^+_\nu$
can no longer mix together to form eigenvectors, but should be
eigenvectors separately. The mixture becomes impossible if and only
if $\Lambda_3$ disappears from (\ref{b23}), (\ref{kappa23}). Then,
according to (\ref{ir1}), $\flat_\nu^{(2)}$ becomes $\sim c^+_\nu$,
hence the degeneracy is expressed as the relation
\bee\label{degen2}\varkappa_1=\varkappa_2,\quad {\rm when}\;
k_0^2-k_3^2=0,\eend accompanied by the
relations\bee\label{Lambda14}\Lambda_3=0,\qquad
-\Lambda_4(k_3^2-k_0^2)= k^2\Lambda_1,\quad {\rm when}\;
k_0^2-k_3^2=0.\eend   Now we use the fact that the electric fields
in eigenmodes are defined up to a common factor to renomalize them
all according to ${\bf\widetilde{e}}={\bf e}(k_0^2-k_3^2)^{1/2}$.
Then, under the special kinematic condition under consideration
$k_0^2-k_3^2=0$, the electric fields in modes 1 and 2 (see eqs.
(\ref{e1}), (\ref{e+}))
 are finite and
equal in length $|{\bf\widetilde{e}}^{(1)}|=
|{\bf\widetilde{e}}^+|=k_\perp k_3$, while that in mode 3 (\ref{e-})
disappears, ${\bf\widetilde{e}}^-=0$. Consequently this degree of
freedom is impossible. As pointed in Section III, in the exceptional
point $k_0^2-k_3^2=0$, the dispersion law $k^2=0$ may correspond to
an actual mode-1 photon. In view of (\ref{degen2}) it must be
accompanied by a mode-2 partner. For $k^2=0$ the electric fields in
modes 1 and 2, besides the fact that they are equal in size, become
polarized in transverse directions, $({\bf\widetilde{e}}^+
{\bf\widetilde{e}}^{(1)})= |{\bf k}_\perp|k_3k_0\sqrt{k^2}=0$.
Therefore, we have again two photon degrees of freedom.

The symmetry relations (\ref{degen1})--(\ref{Lambda14}) are
confirmed in the infrared limit by eqs. (\ref{ir1}) -- (\ref{ir5})
and --  for any value $k_\mu$ of the momentum -- by the one-loop
calculations in QED of \cite{batalin}. In a theory with the dual
invariance, which is not QED, another degeneracy is possible in the
one-invariant case $\mathfrak{G}=0$ \cite{arXiv} that equates the
eigenvalues $\varkappa_2$ and $\varkappa_3,$ since under the
continuous duality transformation the vector
$\left.\flat_\mu^{(2)}\right|_{\mathfrak{G}=0}=(\tilde{F}k)_\mu$ and
the pseudovector $
\left.\flat_\mu^{(3)}\right|_{\mathfrak{G}=0}=(Fk)_\mu$ transform
through each other.
\section{Conclusions} In this paper we studied on the most general
basis  properties of small perturbations of the vacuum, filled with
a constant and homogeneous background electromagnetic field with its
both invariants different from zero. To this end the eigenvector
decomposition of the polarization operator with contribution of
three modes was exploited. We saw how the eigenvectors
characteristic of the one-invariant (magnetic in a special frame)
background field are linearly combined with the help of
dynamics-dependent coefficients to form  eigenvectors of the general
problem under investigation.

Among the vacuum perturbations special attention was payed to the
sourceless excitations that supply poles to the photon propagator
and satisfy three different dispersion equations. These may be
either massive or massless. In the latter case they are called
photons. The massless excitations belong only to two modes, in
accordance with two polarization degrees of freedom of a gauge
vector particle, the photon. Massive excitations belong to all the
three modes, since a massive vector field has three degrees of
freedom. These may have unrestricted number of branches in each mode
depending on the properties of the corresponding dispersion
equation. We described admitted disposition of various dispersion
curves (in the appropriate momentum plane) as it is restricted by
the causal propagation requirement. The eigenmodes are
plane-polarized, and the orientations of their electric and magnetic
fields  with respect to propagation direction and the direction of
the background field are described. We dwelled on the impact the
admixture of an electric field to a magnetic background may have on
the selection rules for photon splitting. We noted that such
admixture results in a larger separation between two different
dispersion curves enhancing the birefringence.

Among possible perturbations of the background caused by small
sources we especially considered the magnetic (part of the) field
produced by a point static electric charge and found its behavior
far from the source.

We also established coincidences between eigenvalues of the
polarization operator (degeneracies) that, under special relations
between momenta, reflect the residual rotational and Lorentz
symmetries of the vacuum left after the background field is imposed.

All the results are approximation-independent, except for the
statement, based on the one-loop calculations, that the mixing
between modes is not resonant in the limit of  small electric field
at the first threshold of electron-positron pair creation by a
photon.

\section*{Acknowledgements}
  This work was supported by the Russian Foundation for
Basic Research (Project No. 05-02-17217) and the President of Russia
Programme (No. LSS-4401.2006.2), as well as by the Israel Science
Foundation of the Israel Academy of Sciences and Humanities.

\section*{Appendix 1}
Suppose, the polarization operator $\Pi_{\mu\nu}$ is known in
components. Then the invariant functions $\Lambda_i, i=1,2,3,4$
involved in its eigenvalues and eigenvectors are given following the
receipts in  \cite{batalin, shabtrudy} as
\bee\label{lambdai}\Lambda_1=d^{(1)}_\mu\Pi_{\mu\nu}(d^{(1)}+d^{(2)})_\nu,\quad
\Lambda_2=- c^{(1)}_\mu\Pi_{\mu\nu}c_\nu^{-},\quad
\Lambda_3=-c^-_\mu\Pi_{\mu\nu}c_\nu^{(3)},\quad
\Lambda_4=-~c^{(3)}_\mu\Pi_{\mu\nu}c_\nu^{+},\quad \eend where
$c^\pm_\nu$ are given by (\ref{d}), and \bee\label{d1} d_\mu^{(1)}=
\frac{\mathcal{E}^2k_\mu-\rmi\mathcal{B}(Fk)_\mu-(F^2k)_\mu-\rmi\mathcal{E}(\tilde{F}k)_\mu}
{2^{1/2}(\mathcal{B}^2+\mathcal{E}^2)^{1/2}(k^2\mathcal{E}^2-kF^2k)^{1/2}}
\Leftrightarrow\frac{{E}^2k_\mu-\rmi{B}(Fk)_\mu-(F^2k)_\mu-\rmi{E}(\tilde{F}k)_\mu}
{2^{1/2}({B}^2+{E}^2)|k_\perp|},\quad\eend

\bee\label{d1+2} d_\mu^{(1)}+d_\mu^{(2)}=2^{1/2}
\frac{\mathcal{E}^2k_\mu-(F^2k)_\mu}
{(\mathcal{B}^2+\mathcal{E}^2)^{1/2}(k^2\mathcal{E}^2-kF^2k)^{1/2}}
\Leftrightarrow 2^{1/2}\frac{{E}^2k_\mu-(F^2k)_\mu}
{({B}^2+{E}^2)|k_\perp|},\eend \bee\label{d3} c_\mu^{(3)}=
\rmi\frac{\mathcal{B}^2k_\mu+\mathcal{E}(Fk)_\mu+(F^2k)_\mu-\mathcal{B}(\tilde{F}k)_\mu}
{(\mathcal{B}^2+\mathcal{E}^2)^{1/2}(k^2\mathcal{B}^2+kF^2k)^{1/2}}\Leftrightarrow
\frac{B^2k_\mu+E(Fk)_\mu+(F^2k)_\mu-B(\tilde{F}k)_\mu}
{(B^2+E^2)(k_0^2-k_3^2)^{1/2}}\eend The notations used here are
connected with those of Refs. \cite{batalin, shabtrudy} as
follows\bee\label{connection}c^{(1,3)}_\mu=\rmi
\sqrt{2}d^{(1,3)}_\mu,\quad c^{-}_\mu
=\rmi(d_\mu^{(1)}-d_\mu^{(2)})/\sqrt{2},\quad c^{+}_\mu
=\rmi(d_\mu^{(3)}-d_\mu^{(4)})/\sqrt{2}.\eend Using the
orthogonality of the polarization operator to vector $k_\mu$ and the
orthogonality of the vector $(F^2k)_\mu$ to the hyperplane spanned
by the two vectors $(Fk)_\mu$ and $(\tilde{F}k)_\mu$, and also the
diagonal representation (\ref{pimunu}), we find that many components
of the vectors between which $\Pi_{\mu\nu}$ is sandwiched disappear
from (\ref{lambdai}). Then we get the simpler representations
(\ref{sandwich}) for the $\Lambda$'s.
 Equation (\ref{sandwich}) for $\Lambda_1$
agrees with eq. (\ref{kappafirst}) and with the relation
$\varkappa_1=\flat_\mu^{(1)}\Pi_{\mu\nu}\flat_\nu^{(1)}/(\flat^{(1)})^2$
that follows from (\ref{eigen}), taking into account the length of
the eigenvector $\flat_\mu^{(1)}$ given in (\ref{first}). Equations
(\ref{sandwich}) for $\Lambda_{2,3,4}$ agree with eqs.
(\ref{eigen}), (\ref{b23}), (\ref{kappa23}), but cannot be deduced
from them, because the latter are  invariant under a similarity
transformation that changes the components of $\Pi_{\mu\nu}$.

 The linear
combinations $\Lambda_i$ of the polarization operator components are
calculated in \cite{batalin} in one-loop approximation, the
calculational details being presented in \cite{shabtrudy} on the
basis of \cite{preprint}. The latter reference as well as
\cite{bayer, gies} contains also calculations
of alternative sets 
of four scalar coefficient functions  of an appropriate set of basic
matrices in terms of which the polarization operator may be
expressed. However, the set of functions $\Lambda_i$ is preferred to
them all, because the eigenvalues are given the simplest in their
terms.

\section*{Appendix 2}

In this Appendix we write the linear in electric field correction
into invariant function $\Lambda_3$ for the case, where an electric
field, much smaller than the magnetic field,  is added parallel to
the latter (this wording refers to the special Lorentz frame). The
linear part of $\Lambda_3$ defines the leading contribution into the
mixing of photon eigenmodes in a  magnetic field due to the
perturbation introduced by the electric field. We present it here as
a result of one-loop calculations of quantum electrodynamics in
external magnetic field.

Using (\ref{d}) and notations (\ref{vectors}) the expansion
(\ref{be}) of the slightly perturbed eigenvectors over the
eigenvectors in a magnetic field alone
becomes\bee\label{pertured}\flat_\mu^{(2)}=-\frac{(\Lambda_2-\Lambda_4)}
{(k^2\mathcal{B}^2+kF^2k)^{1/2}}\left.\flat_\mu^{(2)}\right|_{\mathfrak{G}=0}+
\left(\frac{\mathfrak{G}(\Lambda_2-\Lambda_4)}
{\mathcal{B}^2(k^2\mathcal{B}^2+
kF^2k)^{1/2}}-\frac{\Lambda_3}{(-kF^2k)^{1/2}}
\right)\left.\flat_\mu^{(3)}\right|_{\mathfrak{G}=0},\nonumber\\
\flat_\mu^{(3)}=\frac{(\Lambda_2-\Lambda_4)}
{(-kF^2k)^{1/2}}\left.\flat_\mu^{(3)}\right|_{\mathfrak{G}=0}+
\left(\frac{\mathfrak{G}(\Lambda_2-\Lambda_4)} {\mathcal{B}^2(-
kF^2k)^{\frac1{2}}}-\frac{\Lambda_3
}{(k^2\mathcal{B}^2+kF^2k)^{\frac1{2}}}
\right)\left.\flat_\mu^{(2)}\right|_{\mathfrak{G}=0}. \eend It is
understood that $\Lambda_2-\Lambda_4$ 
in the right-hand side are taken at $\mathfrak{G}=0$. In writing
these equations we took into account that $\Lambda_{2,4}$ are even,
and $\Lambda_3$ is an odd functions of  $\mathfrak{G}.$

It is seen from (\ref{kappa23}) that at $\mathfrak{G}=0$ the
quantity  $-\Lambda_4$ is the polarization operator eigenvalue
$\varkappa_2$ in a magnetic field. Analogously,
$-\Lambda_2=\varkappa_3.$ These quantities in the one-loop
approximation are known \cite{batalin,annphys}. Now we shall write
$\Lambda_3$ in the same approximation by calculating the
$\mathfrak{G}=0$ limit of the corresponding expression from
\cite{batalin}. \bee\label{Lambda5}\Lambda_3=-\frac{\alpha}{4\pi}
\frac{\mathfrak{G}}{\mathfrak{F}}\left(\frac{-kF^2k}{2\mathfrak{F}}\right)^{\frac1{2}}
\left(\frac{-k\tilde{F}^2k}{2\mathfrak{F}}\right)^{\frac1{2}}
\int_0^\infty\frac{\tau\rmd \tau}{\sinh^2
\tau}\int_{-1}^1\rmd\eta(1-\eta)^2\sinh^2\frac{\tau(1+\eta)}{2}\times
\nonumber\\\nonumber\\\times\exp\left\{\frac{kF^2k}{2\mathfrak{F}}\;
\frac{\sinh((1+\eta)/2)\tau)\sinh((1-\eta)\tau/2)}{\sinh(ef\tau)}
-\frac{k\tilde{F}^2k}{2\mathfrak{F}}\;\frac{1-\eta^2}{4ef}\tau-\frac{
m_\rme^2\tau}{ef}\right\}.\qquad\eend Here $\alpha=1/137$ is the
fine structure constant, $m_\rme$ and $e$ are the electron mass and
charge, and $f=\sqrt{2\mathfrak{F}}.$ Note that this expression
vanishes, indeed, either if
$-\frac{k{F}^2k}{2\mathfrak{F}}\Leftrightarrow k_\perp^2=0,$ or if
$-\frac{k\tilde{F}^2k}{2\mathfrak{F}}\Leftrightarrow k_0^2-k_3^2=0,$
as it should in accordance with what the symmetry of the external
field prescribes, as it was established in the body of this article,
eqs. (\ref{degen1}), (\ref{Lambda14}). The integral (\ref{Lambda5})
has an infinite number of branching points, the same as
$\Lambda_{1,2,4}$, with singular inverse square-root behavior - the
cyclotronic resonances at thresholds of electron-positron pair
creation by a photon \cite{annphys, shabtrudy}. However, the
lowest-lying resonance is not in the point (in the  variables,
referring to the special frame) $k_0^2-k_3^2=4m_\rme^2$, like in
$\varkappa_2=-\Lambda_4$, but in the point
$k_0^2-k_3^2=[(m_\rme^2+2ef)^{1/2}+m_\rme]^2,$ like in
$\varkappa_3=-\Lambda_2$, because this value  borders  the
convergence domain of the $\tau$-integration in (\ref{Lambda5}).
This means that the (small as compared with the magnetic) electric
field cannot affect the phenomenon of the mode-2 photon capture with
its adiabatic conversion into a free \cite{nature} or bound
\cite{ass} electron-positron pair in the lowest Landau level.


\begin{thebibliography}{40}
\bibitem{footnote3}The situation where
they merge \cite{plebanski} to become one curve -- to be more
precise, one straight line -- is exemplified  by  Born-Infeld
nonlinear electrodynamics. In that case the square root in
(\ref{birefringence}) vanishes on the common mass shell of modes 2
and 3.
\bibitem{reservation}With the reservation that mode 1 was ignored
 in \cite{adler}.
\end{thebibliography}
\end{document}